# A review of theories and models utilized by empirical studies about mental health help-seeking and implications for future research

**Jiaying "Lizzy" Liu** University of Texas at Austin School of Information
**Yongjie Sha** University of Michigan School of Information
**Yan Zhang** University of Texas at Austin School of Information

**Abstract:**
Purpose: With the rise of mental health risks globally, it is urgent to provide effective mental health support. However, a holistic understanding of how people seek help for mental health problems remains limited, impeding the development of evidence-based intervention programs to facilitate help-seeking behavior. This study reviews current theories that guide empirical research on young adults' help-seeking behavior using technologies, identifies limitations in existing frameworks, and proposes directions for future research.
Methods: We searched databases that are most likely to contain mental health help-seeking practices with relation to information technology, including PubMed, ACM Digital Library, Web of Science, PsycInfo, ScienceDirect, EBSCO, and Cochrane Library.
Results: Of 2443 abstracts reviewed, 43 studies met the criteria and were included in the analysis. We identified 16 theories and models. They represent seven perspectives to view mental health help-seeking and reveal factors such as accessibility, stigma, and social support as key factors influencing help-seeking.
Limitations: We summarized the theories and models and categorized them based on their primary perspective. Cross-perspective connections could be explored in future reviews.
Conclusions: A holistic approach to creating culturally sensitive multi-level interventions that consider individual, interpersonal, and community factors is needed to advance effective mental health help-seeking support strategies.

## 1. Introduction

One in five US adults show symptoms of mental illness in 2021 (1). Mental illnesses, ranging from depression and bipolar disorder to schizophrenia, vary in severity but collectively pose a significant public health challenge. Mental health help-seeking refers to an adaptive coping mechanism that involves the proactive pursuit of external support to address psychological issues (2). Effective help-seeking is paramount for managing and recovering from mental health disorders. It provides access to professional guidance (3), supportive networks (4), and necessary treatment modalities (5), ultimately promoting overall well-being (6).

However, only a limited percentage of individuals with mental health problems actually seek help (7). For instance, a UK national survey revealed that approximately 40% of people with bipolar disorder did not obtain mental health care in 2013 (8). Individuals seek support from various sources, including mental health professionals, service providers, informal social networks, and increasingly digital platforms (9). The growing integration of technology into healthcare has amplified the role of technologies in facilitating help-seeking (10).

Previous reviews have identified numerous barriers to mental health help-seeking, including stigma, negative beliefs about professional services (11), sociocultural norms (12), lack of mental health literacy, and limited availability of professional help (13; 14). However, these reviews have largely neglected the theoretical foundations for designing interventions, which are important for understanding mechanisms that support successful help-seeking.

To bridge this gap and inform future research, we reviewed models and theories used to guide mental health help-seeking interventions in published empirical studies. This review aims to address two research questions: What factors influence people's mental health help-seeking? What implications do theories about mental health help-seeking have for future intervention studies?

## 2. Methods

We summarized the literature search and screening process in Figure 1.

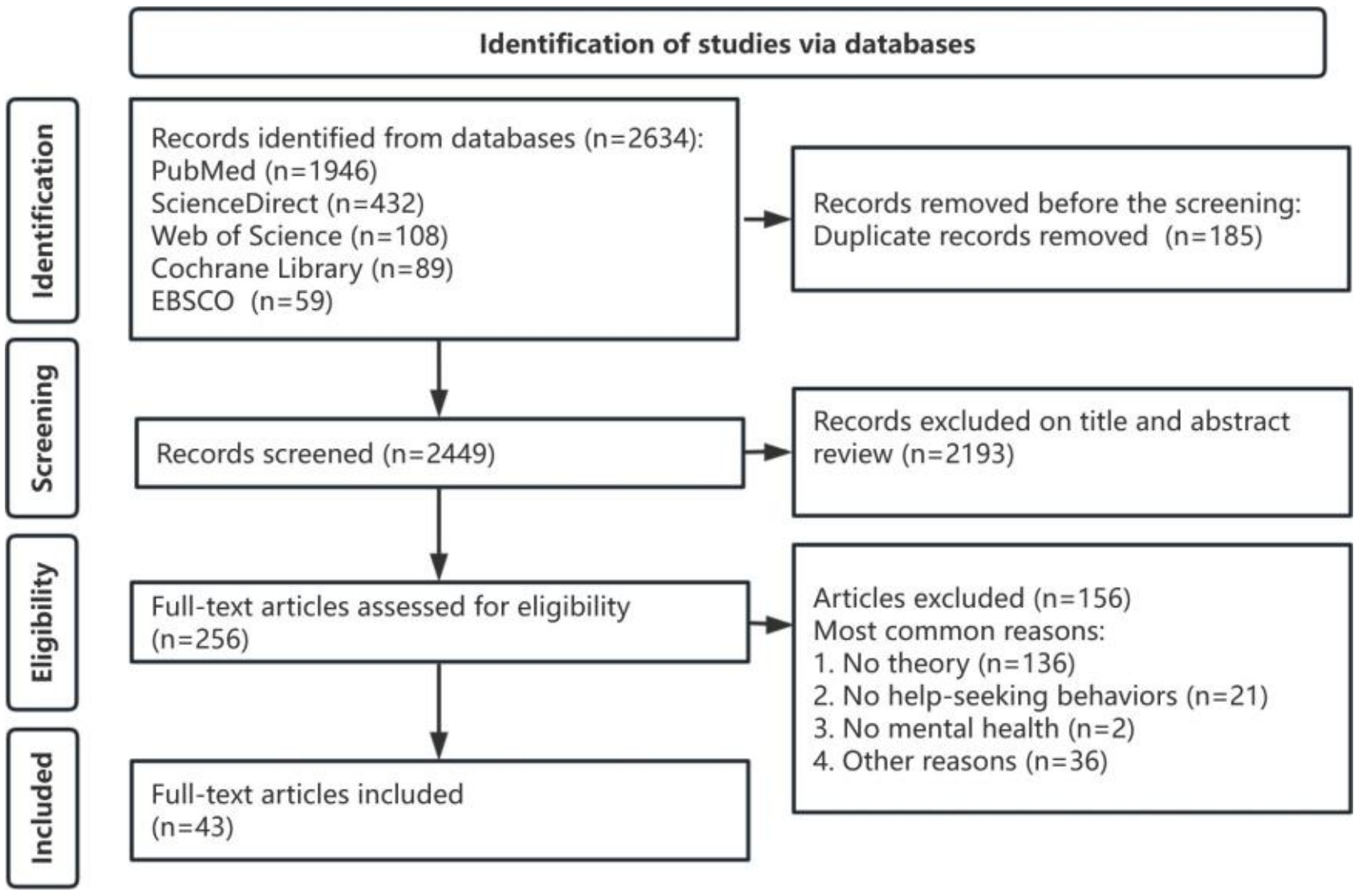

**Figure 1:** Searching and Screening Process.

## 2.1. Literature Search

We searched the databases that were most likely to contain mental health help-seeking practices involving information technology, including PubMed, ACM Digital Library, Web of Science, PsycInfo, ScienceDirect, EBSCO, and Cochrane Library, in February 2023. The search query was (mental health OR mental illness OR mental disorder) AND (help seeking OR management) AND (technology OR social media OR online forum OR bot OR app OR track OR Instagram OR Facebook OR Twitter OR reddit OR internet OR website OR chatbot OR robot) and limited to the title, abstract, and keyword fields. The initial search returned 2634 articles. After removing 185 duplicates, 2443 unique articles were retained for screening.

## 2.2. Inclusion and Exclusion Criteria, and Screening

We included articles that applied existing theories or models to guide their research design or to describe or explain mental health help-seeking. Articles were included if they (1) are empirical studies, (2) focus on mental health help-seeking behaviors such as self-management, self-tracking, information-seeking, and seeking social support, (3) involve theories or models for mental health help-seeking research (4) were peer-reviewed, and (5) were written in English. Literature reviews or perspective papers were excluded. We screened 2443 unique articles with the assistance of Rayyan.ai, a website that supports collaborative literature screening. Two authors first independently assessed 150 articles based on the title and abstract as include, exclude, or maybe. They then discussed the conflicting ratings with the third author. This process clarified the inclusion and exclusion criteria. The two authors then reviewed the remaining articles based on the title and abstract and discussed maybe articles in weekly meetings with the third author. The process resulted in 256 articles for full-text review. After the review, the final sample included 43 articles.

## 2.3. Data Extraction

We utilized Microsoft Excel to facilitate the analysis. We extracted theories/models utilized, variables in them, and associated empirical research results. We also retrieved the original publications of these theories and models and extracted elements, including premises, main ideas, concepts, and assumptions, to better understand them.

Subsequently, we categorized theories into six categories based on their perspectives. The three authors held weekly meetings to discuss the categorization. In instances where a theory/model encompassed elements fitting multiple categories, we assigned it to the category based on its primary perspective.

## 3. Results

We identified 16 theories and models from the included 43 empirical studies on mental health help-seeking. These theories and models were categorized into seven categories, detailed in Table 1.

Table 1. Categories of theories/models applied in empirical studies about mental health help-seeking behaviors

| Theory / Model | Summary | Factors in the theory/model | Examples |
|---|---|---|---|
| ***Category 1: Psychological Perspective (3 theories and 7 empirical studies)*** | | | |
| Cognitive Theory of Depression / The negative triad (15) | Depressed individuals have a persistent negative schema, biasing thoughts toward the self, environment, and future. As depression worsens, negative automatic thoughts and biases become more frequent and pervasive. | -Self-distancing<br>-Self-compassion | (16) (17) (18) |
| The mental health continuum model (19) | Mental health is not just the absence of illness but a combination of positive feelings and functioning. Flourishing represents good mental health, while its absence is called languishing. | -Resilience<br>-Stigma and discrimination | (20) |
| Broaden-and-Build Theory (21) | Positive emotions broaden a person's mindset and help build lasting resources. Infusing these emotions can trigger increased help-seeking if opportunities are made clear and available during that time.<br>-Confidence to seek mental health information. | -Positive emotions infusion (elevation, gratitude, savoring)<br>-Dispositional hope | (22) (23) (24) |
| ***Category 2: Motivation Perspective (2 theories and 11 empirical studies)*** | | | |
| Theory of planned behavior/reasoned action (25) | Behavior is driven by the intention to act, which depends on attitude, perceived social norms, and self-control. | -Intention<br>-Attitude<br>-Subjective norms (e.g., self-stigma, negative help-seeking attitudes)<br>-Self-efficacy (e.g., confidence accessing mental health information) | (26) (27) (28) (29) (30) (31) (32) |
| Self-determination theory (SDT) (33) | This macro theory of motivation and personality explores people's innate growth tendencies and psychological needs, focusing on self-motivation and self-determination in the absence of external influences. | -Autonomy<br>- Competence<br>- Being connected/Relatedness | (34) (35) (33) (36) |

***Category 3: Service Access Perspective (2 theories and 6 empirical studies)***

| | | | |
|---|---|---|---|
| Anderson's behavioral model of health services use (R. Andersen, 1968) | Health service use and personal health practices are predicted by predisposing factors, resources, perceived need, and the health care system. | -Personal and community enabling resources (e.g., income, health insurance, regular source of care, travel and waiting times, social relationships, informal resources, community resources, formal mental health care resources)<br>-Perceived need (e.g., health education programs) | (37) (38) (39) |
| The pathways to psychiatric care (40) | The five-layer referral system outlines care pathways, starting with general practitioners and moving through primary care, psychiatric care, and hospital admission. | -Attitudinal barriers (e.g., control and self-reliance, minimizing problems, sense of resignation that nothing will help)<br>-Social environment (college experience with more open conversations)<br>-Psychological and physical distance from family<br>-Culture (traditional beliefs, hiding as strength)<br>-Self-reliance (impact on informal HS intentions mediated by perceived social support)<br>-Workplace climate (perceived social support) | (41) (42) (43) |

***Category 4: Identity Perspective (3 theories and 6 empirical studies)***

| | | | |
|---|---|---|---|
| Positioning theory (44) | Individual subjectivity is shaped by available positions in discourse, which are sites of acceptance, resistance, or negotiation. People reinforce positive identities within broader norms while resisting stigmatized mental illness positions. | -Stigma<br>-Social identification | (45) |
| Social identity perspective (SIP) (46) | Experiencing discrimination can make a stigmatized group central to one's social identity, shaping attitudes and behavior. This identity can link mental health literacy to personal experience and encourage future help-seeking. | -Stigma<br>-Social identification<br>-Perceptions about the group<br>-Self-relevant step<br>-Service utilization | (47) (48) (49) |
| Identity threat model of stigma (50) | Situational cues, stigma perceptions, and personal beliefs shape how stigma is appraised. Identity threat triggers stress and coping efforts, which impact health outcomes. | -Perceived public stigma (negative attitudes held by others)<br>-Personal stigma (negative attitudes toward oneself)<br>-Trust in health professionals<br>-Community attitudes towards mental illness<br>-Internalized masculine norms.<br>-Familiarity with mental illness/literacy<br>-Social distance<br>-Beliefs about the causes of mental illness<br>-'Silencing' of mental illness resulting from heightened levels of ideological stigma<br>-Stigma (re)production and maintenance at community level | (51) (52) |
| ***Category 5: Illness Journey Perspective (2 theories and 5 empirical studies)*** | | | |
| Cycle of Avoidance (COA) model (53) | The model highlights tensions in recognizing, accepting, and avoiding mental distress. Key actions that delay help-seeking include normalizing symptoms, offering alternative explanations, accommodating severe distress, raising the threshold for "real" distress, and postponing professional help. Similar strategies include relying on self-resilience, denying the benefits of help, and questioning help-seeking. | -Lay diagnosis<br>-Coping and normalization strategies<br>-Meaning of help and help-seeking | (54) (55) |
| Stages of help-seeking decision-making model (2; 56) | The four stages of help-seeking are: (1) recognizing the problem, (2) expressing the need for support, (3) knowing available help sources, and (4) being willing to share personal information. | -Accessibility of formal services<br>-Social factors (system navigation)<br>-Health literacy (symptom recognition, acting on symptoms)<br>-Low-intensity sources of support such as self-help<br>-Perception of access (cost and financial barriers, adequacy of care, timeliness)<br>-Stigma attitudes based on social distance and willingness to disclose mental health issues to campus members<br>-Self-labelling (impact of personal stigma) | (57) (58) (59) |
| ***Category 6: Sociocultural Perspective (3 theories and 7 empirical studies)*** | | | |
| Network episode model (60) | Help-seeking is a social process shaped by one's community, treatment systems, and social service agencies. | -Social network (e.g., component, structure)<br>-Help-seeking contacts (first contact, types of contacts, frequency)<br>-Mental health literacy<br>-Use of online resources<br>-Family members' support (e.g., parental authoritativeness, help-seeking skills, help-seeking knowledge<br>-Acculturation | (61) (62) |
| Cultural determinants of help-seeking (63) | Designed for cross-cultural use, the model examines social and cultural barriers and facilitators of help-seeking. Cultural determinants are internal and social factors that influence help-seeking. | -Illness interpretation (causal beliefs, sense of coherence, attitudes toward treatment)<br>-Social context (social support, social conflict)<br>-Illness consequence (beliefs about mental illness)<br>-Perceived need for mental health care | (64) (65) |
| Community mental health competency framework (CMHC) (66) | Communities can promote, prevent, and treat mental health while advocating for it. CMHC programs must address stigma and, low mental health awareness, advocate for responsive services, and tackle social factors like gender inequality and poverty. | -Cultural competence and knowledge<br>-Social networks and participation<br>-Social context | (67) (68) (69) |
| ***Category 7: Sociotechnical Perspective (1 theory and 1 empirical study)*** | | | |
| Sociotechnical ecology of serious mental illness management (70) | The ecology of personal data practices supports self-management for individuals with mental health concerns. It includes micro, exo, and info layers, plus a cross-cutting temporal layer. Relational interactions are defined by four dimensions: valence, intensity, directionality, and dynamism. | -Personal informatics practices<br>-Social relations and roles of healthcare professionals, inner social circles, outer circles and sociocultural backdrops<br>-Interpersonal engagement through personal informatics | (71) |

### 3.1. Psychological perspective

Three theories/models trace the psychological foundation of mental health conditions to explain how individuals' cognitive and emotional capacity influences their ability and intention to seek help. For example, the Cognitive Theory of Depression (15) explains why people with depression resist help-seeking, positing that negative cognitive schemas hinder behavioral change by influencing one's perceptions of themselves, the world, and the future. The Broaden-and-Build Theory (21) goes beyond cognitive factors to underscore the role of emotions, suggesting that positive emotions such as joy and gratitude can broaden an individual's momentary thought-action repertoire and cultivate resilience, which facilitates further help-seeking.

These theories have guided interventions aimed at modifying individuals' cognition about mental health and emotional states to encourage help-seeking behaviors and psychological flourishing (22). For example, interventions teach individuals strategies such as self-distancing (16) and self-compassion (17). Additionally, researchers have designed interventions for savoring positive experiences and emotions (24). Stories and narratives have also been employed to boost emotion elevation and increase help-seeking intentions (23).

### 3.2. Motivation perspective

Two theories and models explore individuals' motivations to seek help. For example, the theory of planned behavior (25) suggests that people seek help to gain autonomy, agency, and self-awareness. Similarly, self-determination theory (33) views help-seeking as an assertion of self-determination, agency, and control over one's own life.

Individuals' motivation to seek mental health help is influenced by multiple factors. The theory of planned behavior demonstrates that help-seeking behavior is shaped by attitudes, subjective norms, and perceived behavioral control (25). Studies found that emotional factors also play a crucial role (30) and influence individuals' selection of help sources (72). For instance, individuals often prefer seeking help from family members rather than medical professionals due to more supportive responses (26). This preference may be attributed to the reduced social distance with family members, which correlates with increased contact and help-seeking intentions (28).

### 3.3. Service access perspective

Two theories and models specify how individuals seek and utilize mental health services through distinct pathways. The Goldberg–Huxley model (40) details the pathway of people accessing psychiatric care, with particular emphasis on "points of first contact" (73) as crucial pathway initiators that shape service perceptions, future choices, and expectations. Subsequent studies explored various "points of first contact," such as non-traditional sources like native or religious healers and even police institutions (74). For example, Pendse et al. (75) emphasized the role of the national helpline as a crucial first point of contact, underscoring its impact on individuals' subsequent engagement with mental health care resources. Several studies indicate that the most common preferences for support were internet-based information resources, followed by support provided by general practitioners and maternal child health nurses (41).

Anderson's model, which views help-seeking as a form of service use, explains health service utilization through three key factor clusters: predisposing, enabling, and need factors (76; 77; 78). Empirical studies have consistently supported the role of these clusters in shaping help-seeking behavior. Predisposing factors, such as age and education, have been shown to influence individuals' attitudes and readiness to seek care, often acting as barriers (41). Need-related factors, including self-reliance, self-neglect severity, and symptom resignation, reflect the perceived urgency or necessity for mental health services and have been validated as critical drivers of help-seeking (41; 79). Enabling factors, such as social support, health literacy, and neighborhood communicativeness, play a practical role by either facilitating or hindering access to care. Later research has built on this foundation, identifying additional factors that refine the model. For example, Suka et al. (37) introduced population-specific contextual factors, while Lu et al. (80) include detailed characteristics across biological (e.g., age, gender), clinical (e.g., symptom severity), behavioral (e.g., drug/alcohol use), and psychological (e.g., internal assets) domains.

### 3.4. Identity threat perspective

Three theoretical frameworks, including positioning theory (44), social identity theory (81), and the identity threat model of stigma (50), examine the tensions between seeking external help and maintaining self-identity. These theories suggest that interactions with healthcare institutions can position individuals as "patients" (44), potentially threatening their sense of self when this categorization deviates from their normal identity. Stigma thus emerges as a central concept from this perspective, with numerous empirical studies investigating how perceived public stigma and personal stigma manifest across various contexts (51). Research indicates that higher levels of stigma correlate with increased identification with stigmatized groups, subsequently decreasing intentions and behaviors of healthcare service utilization (48).

Acknowledging that identity is formed and enacted through social interactions (81), these models emphasize how perceptions of social positions shape patterns of resource utilization during help-seeking. Group identification—the positioning of oneself within a particular social group—emerges as a significant predictor of help-seeking attitudes, even after accounting for established predictors (47). Help received from ingroup sources is more readily accepted and poses less threat to one's identity compared to assistance from perceived outgroup sources (47).

### 3.5. Illness journey perspective

Two theories are from illness journey perspective and focus on the real-world help-seeking process and stages through which individuals move from non-help-seeking to help-seeking (53). For example, Rickwood et al. (2)'s help-seeking model specifies four stages of the help-seeking journey: recognition, decision-making of help-seeking, help-seeking actions, and continuation and maintenance. Researchers frequently use this model to guide the identification of barriers and facilitators to mental health help-seeking. For example, many find the first stage of recognizing mental illness critical. Research indicates that adolescents are more likely to seek help from professionals and friends when they can recognize mental illness (57). Being aware of mental illness serves as a coping mechanism for some individuals, increasing their likelihood of seeking help from health service providers (58). Across the stages, studies found that non-professional help-seeking is more prevalent in earlier stages of the journey, while professional help-seeking typically occurs in later stages (9).

Viewing help-seeking as a journey enables a deeper theoretical understanding of how complex factors collectively shape individuals' real-world help-seeking experiences. For instance, Biddle et al. (53)'s cycle of avoidance model highlights the dynamic process where individuals constantly negotiate between help-seeking and non-help-seeking, impacted by concerns of stigmatization, solitary feelings, and resource availability. Centering on people's subjective experiences, researchers have frequently used qualitative approaches to uncover contextual barriers in mental health help-seeking experienced by specific populations such as young adults (54). Studies commonly reveal tensions around making sense of mental health concerns (82), avoiding mental distress (83), and combating perceived stigmas (84; 55).

### 3.6. Sociocultural perspective

Three theories and models view mental health help-seeking from a sociocultural perspective. They respectively highlight the role of social networks (85), community environments (86), and cultural beliefs (63), in shaping help-seeking behaviors. Studies deploying theories from this perspective investigate how the specific content, structure, and function of family networks, community systems, educational institutions, and treatment providers influence individuals' pathways to care (62). Research has revealed that personal support networks can sometimes raise stigma concerns (87), particularly in certain cultural contexts like Hispanic communities, where mental health stigma tends to be more pronounced (64; 88). Similarly, Ohta et al. (65) found community attitudes toward mental illness significantly predict help-seeking behaviors.

Studies use sociocultural theories to guide mental health interventions. Community-facilitated programs have demonstrated the potential to support sustained help-seeking behavior by tapping into existing social structures and cultural frameworks (89; 61). These programs have been particularly effective in supporting individuals without personal support networks, especially among marginalized populations such as immigrants and older adults. For example, Mathias et al. (67) documented a community-based mental health project in India, where researchers organized various community activities around mental health. Through informal conversations, patient role-playing exercises, and group participation, the project promoted social support and inclusion. By instigating new knowledge into existing social frameworks, the initiative increased community understanding and created safer social environments. Mannel et al. (68) identified how community-focused policy interventions, such as involving police, social services, and legal support, can streamline access to resources and foster community support through active engagement in discussions.

### 3.7. Sociotechnical Perspective

Complementing the sociocultural perspective, the sociotechnical ecology of serious mental illness management (70) highlights an integrated social and technological perspective as technologies become an integral part of the mental health resource infrastructure. It is important to note the user-centered design and the interplay of barriers and facilitators in managing mental health through the interface design and technical affordances of technology-mediated mental health services, such as self-tracking apps and sensing devices. This holistic framework underscores the need for multidimensional interventions that address the interplay of social and technological factors in promoting mental health and well-being.

## 4. Discussion

We reviewed 16 theories and models in 43 empirical studies that examine people's mental health help-seeking behaviors and identified seven distinct perspectives. Mapping these perspectives across a timeline shows that the conceptualization of mental health help-seeking has evolved over time, as shown in Figure 2.

Early models [1970s-1980s] like the cognitive theory of depression (15), the theory of planned behavior (25), and the behavioral model of health resource use (76), primarily explain help-seeking decisions based on individuals' psycho-biological status, intentions to achieve behavior change goals, and the severity of mental health problems. Later models [1990s-2000s] introduced identity (44) and illness journey (53) concepts, viewed help-seeking as a complex ongoing journey where individuals continuously make sense of their illness and perceive help-seeking behaviors as a threat to their identity. Cultural factors started to appear in early 2000s but were not widely acknowledged as the main influencing factors. More recent theories and models [2010s-present] adopted sociocultural (63) and sociotechnical (70) perspectives, explicitly acknowledging the role of multilevel factors, ranging from personal, interpersonal, community, social, and technological, in shaping help-seeking behaviors.

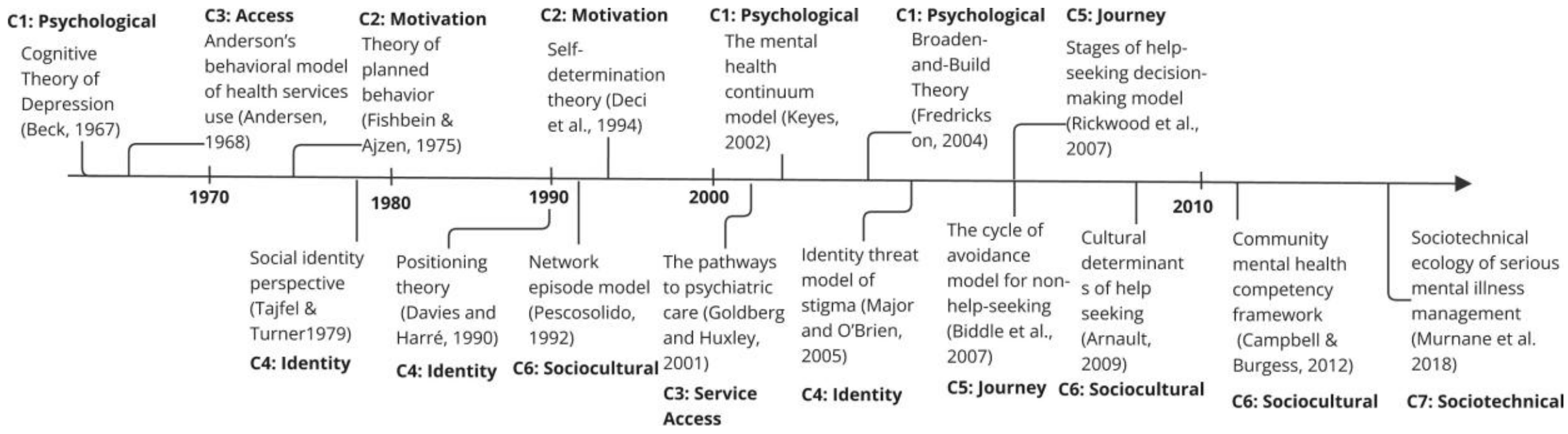

**Figure 2:** Timeline of theories/models from the seven perspectives.

## 4.1. Theoretical integration and theory-driven mental health interventions

Over time, theoretical frameworks have evolved to become increasingly multilayered and systematic. However, significant gaps remain between different theoretical perspectives, highlighting the need for a comprehensive approach that integrates diverse theoretical viewpoints.

Theory integration can lead to more innovative and effective interventions. For instance, combining the motivation perspective's focus on individual education (90) with the sociocultural perspective's emphasis on interpersonal and community support (70) could yield more powerful intervention programs by increasing individuals' mental health literacy, the supportiveness of peer networks, and thus encourage sustained help-seeking behaviors (91; 92). Emerging empirical studies have begun integrating these two approaches, revealing that peer-led mental health literacy programs can both impart knowledge and provide immediate social support (93), thereby addressing multiple facets of the help-seeking process simultaneously.

As another example, both the identity threat perspective and the illness journey perspective can be enriched by incorporating insights from the psychological perspective to illuminate the psychological nuances underlying individuals' behavioral barriers in the help-seeking process. For example, while Rickwood's model (2) delineates four behavioral stages and has been widely adopted in subsequent research (94), it does not fully explain the underlying mechanisms driving progression through these stages. This gap was highlighted by Yamasaki et al. (57), who demonstrated that symptom recognition and mental health illness self-identification are distinct steps, suggesting the need to identify the psychological factors that differentiate these processes.

Moving forward, researchers should explore additional opportunities for theoretical integration to develop comprehensive and ecologically valid frameworks to inform the development of more effective, tailored interventions that can address specific groups' and communities' needs for mental health help-seeking.

## 4.2. Key factors influencing mental health help-seeking behaviors

Across the seven perspectives, three factors emerge as critical influencers of individuals' help-seeking intentions and behaviors: accessibility of mental health resources, stigma surrounding mental health problems, and social support.

Source accessibility is foundational for help-seeking (95). Yet mental health services are unavailable, unaffordable, or geographically distant for many populations, especially in low-resource areas such as rural areas and the global south (96). Beyond the physical and financial aspects of accessibility, the reviewed theories amplify other dimensions of accessibility. For example, the illness journey perspective highlights how individuals navigate healthcare systems, underscoring the need for clear pathways to care to enhance accessibility in a real-life context (75). The sociocultural perspective suggests that accessibility is not just about the presence of services but also their cultural relevance and community acceptance (97; 98).

Stigma pervades all levels of the help-seeking process (64; 58). At the individual level, theories and models from the psychological and motivation perspectives demonstrate how internalized stigma (99) can delay symptom recognition and help-seeking. In the interpersonal, community, and social contexts, as reflected in the illness journey, identity threat, and sociocultural perspectives, fear of judgment can inhibit disclosure (50). A multi-level approach to destigmatization, informed by these perspectives, could target individual beliefs, interpersonal interactions, and community narratives.

Social support, less investigated until the last decade, has emerged as a pivotal factor (70; 100). Mounting evidence suggests that social support not only buffers mental health problems but also facilitates help-seeking (101). The sociocultural perspective highlights how supportive others can validate concerns and guide individuals to professional help (92). The ecological perspective further suggests that community-level support, such as mental health advocacy groups, can shift social norms towards acceptance of help-seeking (102).

### 4.3. Implications for digital mental health help-seeking interventions

This review highlights the emerging sociotechnical perspective, which addresses the pervasive integration of technologies into mental health resources. This perspective (70) emphasizes the interplay between social and technological factors in shaping help-seeking behaviors, suggesting that digital mental health interventions should be designed with consideration for broader social, cultural, and environmental contexts.

Future research should explore how digital technologies' unique affordances can enhance mental health help-seeking. Studies indicate that mobile apps offering real-time support and personalized interventions based on user data could address the dynamic nature of mental health needs more effectively than traditional static interventions (103). Human-computer interaction theories, such as the Technology Acceptance Model (104), can inform these digital interventions. User-centered design approaches require meticulous tailoring of tools to diverse user needs, potentially increasing perceived usefulness and ease of use, thereby enhancing the effectiveness and adoption rates of digital mental health interventions across diverse user groups (105).

While the current sociotechnical perspective is still developing, it can benefit from incorporating more traditional perspectives to create digital mental health interventions that are technologically advanced, socially conscious, ethically sound, and effective in supporting individuals' mental health help-seeking journeys (106; 107). For instance, integrating psychological and motivational perspectives, such as positive psychology and emotion regulation strategies suggested by the Broaden-and-Build Theory of Positive Emotions (21), may increase the effects of persuasive technologies in nudging people's help-seeking behaviors (108).

In addition, digital interventions should work in concert with traditional services. Researchers pointed out that these technologies, such as care robots (109) and self-management apps (110) should complement, not replace, existing resources and in-person care systems. Collaborative efforts involving researchers, practitioners, policymakers, and community stakeholders are essential to developing comprehensive interventions that leverage technology while addressing sociocultural factors and structural barriers (9).

### 4.4. Limitations

This review has several limitations. First, it focused on theories and models related to mental health help-seeking. Thus, the empirical results presented in the review were those related to the theories and models. Future reviews could expand on the inclusion of empirical results. Second, while we categorized theories/models based on their primary perspective, many include constructs relevant to multiple perspectives. These cross-perspective connections could be more explicitly explored in future reviews.

## 5. Conclusion

This review contributes to a better understanding of mental health help-seeking by identifying seven theoretical perspectives that have guided empirical research on this behavior and key factors such as accessibility, stigma, and social support that influence the behavior. Developing effective interventions that address individual motivations, interpersonal dynamics, and community contexts requires a holistic theoretical understanding of help-seeking, which calls for the informed integration of multiple theoretical perspectives.

## References

[1] National Institute of Mental Health. Mental Illness; 2023. Available from: https://www.nimh.nih.gov/health/statistics/mental-illness.

[2] Rickwood D, Thomas. Conceptual measurement framework for help-seeking for mental health problems. Psychology Research and Behavior Management. 2012 Dec:173. Available from: http://www.dovepress.com/conceptual-measurement-framework-for-help-seeking-for-mental-health-pr-peer-reviewed-article-PRBM.

[3] Vogel DL, Wester SR, Larson LM, Wade NG. An information-processing model of the decision to seek professional help. Professional Psychology: Research and Practice. 2006 Aug;37(4):398-406. Available from: http://doi.apa.org/getdoi.cfm?doi=10.1037/0735-7028.37.4.398.

[4] Kordy H, Backenstrass M, Hüsing J, Wolf M, Aulich K, Bürgy M, et al. Supportive monitoring and disease management through the internet: An internet-delivered intervention strategy for recurrent depression. Contemporary Clinical Trials. 2013;36(2):327-37. Section: 0. Available from: https://www.sciencedirect.com/science/article/pii/S1551714413001298.

[5] Fonseca A, Gorayeb R, Canavarro MC. Women's use of online resources and acceptance of e-mental health tools during the perinatal period. International Journal of Medical Informatics. 2016;94:228-36. Section: 0. Available from: https://www.sciencedirect.com/science/article/pii/S1386505616301770.

[6] Sacks S, McKendrick K, Sacks JY, Banks S, Harle M. Enhanced outpatient treatment for co-occurring disorders: main outcomes. Journal of Substance Abuse Treatment. 2008 Jan;34(1):48-60.

[7] Stewart G, Kamata A, Miles R, Grandoit E, Mandelbaum F, Quinn C, et al. Predicting mental health help seeking orientations among diverse Undergraduates: An ordinal logistic regression analysis. Journal of affective disorders. 2019;257:271-80.

[8] Humpston CS, Bebbington P, Marwaha S. Bipolar disorder: Prevalence, help-seeking and use of mental health care in England. Findings from the 2014 Adult Psychiatric Morbidity Survey. Journal of affective disorders. 2021;282:426-33.

[9] Liu J, Zhang Y. Exploring Young Adults' Mental Health Help-Seeking Journey: Preliminary Findings on Resource Navigation Behavior. Proceedings of the Association for Information Science and Technology. 2024;61(1):570-5. _eprint: https://onlinelibrary.wiley.com/doi/pdf/10.1002/pra2.1060. Available from: https://onlinelibrary.wiley.com/doi/abs/10.1002/pra2.1060.

[10] Tsirmpas C, Andrikopoulos D, Fatouros P, Eleftheriou G, Anguera JA, Kontoangelos K, et al. Feasibility, engagement, and preliminary clinical outcomes of a digital biodata-driven intervention for anxiety and depression. Frontiers in Digital Health. 2022;4:868970.

[11] Aguirre Velasco A, Cruz ISS, Billings J, Jimenez M, Rowe S. What are the barriers, facilitators and interventions targeting help-seeking behaviours for common mental health problems in adolescents? A systematic review. BMC Psychiatry. 2020 Jun;20(1):293. Available from: https://doi.org/10.1186/s12888-020-02659-0.

[12] Möller-Leimkühler AM. Barriers to help-seeking by men: a review of sociocultural and clinical literature with particular reference to depression. Journal of Affective Disorders. 2002 Sep;71(1):1-9. Available from: https://www.sciencedirect.com/science/article/pii/S0165032701003792.

[13] Goetter EM, Herbert JD, Forman EM, Yuen EK, Gershkovich M, Glassman LH, et al. Delivering exposure and ritual prevention for obsessive–compulsive disorder via videoconference: Clinical considerations and recommendations. Journal of Obsessive-Compulsive and Related Disorders. 2013 Apr;2(2):137-45. Available from: https://www.sciencedirect.com/science/article/pii/S2211364913000043.

[14] Hom MA, Stanley IH, Joiner TE. Evaluating factors and interventions that influence help-seeking and mental health service utilization among suicidal individuals: A review of the literature. Clinical Psychology Review. 2015 Aug;40:28-39.

[15] Beck AT. Depression: Clinical, experimental and theoretical aspects. New York: Harper and Row; 1967.

[16] Hollar SM, Siegel JT. Self-distancing as a path to help-seeking for people with depression. Social Science & Medicine. 2020 Jan;245:112700. Available from: https://www.sciencedirect.com/science/article/pii/S0277953619306951.

[17] Dschaak ZA, Spiker DA, Berney EC, Miller ME, Hammer JH. Collegian help seeking: the role of self-compassion and self-coldness. Journal of Mental Health. 2021 May;30(3):284-91. Publisher: Routledge _eprint: https://doi.org/10.1080/09638237.2019.1677873. Available from: https://doi.org/10.1080/09638237.2019.1677873.

[18] Becker EM, Jensen-Doss A. Computer-Assisted Therapies: Examination of Therapist-Level Barriers to Their Use. The Theory-Practice Gap in Cognitive-Behavior Therapy. 2013 Dec;44(4):614-24. Available from: https://www.sciencedirect.com/science/article/pii/S000578941300049X.

[19] Keyes CLM. The Mental Health Continuum: From Languishing to Flourishing in Life. Journal of Health and Social Behavior. 2002;43(2):207-22. Publisher: [American Sociological Association, Sage Publications, Inc.]. Available from: https://www.jstor.org/stable/3090197.

[20] Fernandez DK, Deane FP, Vella SA. Adolescents' Continuum and Categorical Beliefs, Help-seeking Intentions, and Stigma Towards People Experiencing Depression or Schizophrenia. International Journal of Mental Health and Addiction. 2022 Dec;20(6):3285-300. Available from: https://doi.org/10.1007/s11469-022-00766-5.

[21] Fredrickson BL. The broaden-and-build theory of positive emotions. Philosophical Transactions of the Royal Society B: Biological Sciences. 2004 Sep;359(1449):1367-78. Available from: https://www.ncbi.nlm.nih.gov/pmc/articles/PMC1693418/.

[22] Kaniuka AR, Job SA, Brooks BD, Williams SL. Gratitude and lower suicidal ideation among sexual minority individuals: theoretical mechanisms of the protective role of attention to the positive. The Journal of Positive Psychology. 2021 Nov;16(6):819-30. Publisher: Routledge _eprint: https://doi.org/10.1080/17439760.2020.1818814. Available from: https://doi.org/10.1080/17439760.2020.1818814.

[23] Siegel JT, Thomson AL. Positive emotion infusions of elevation and gratitude: Increasing help-seeking intentions among people with heightened levels of depressive symptomatology. The Journal of Positive Psychology. 2017 Nov;12(6):509-24. Publisher: Routledge _eprint: https://doi.org/10.1080/17439760.2016.1221125. Available from: https://doi.org/10.1080/17439760.2016.1221125.

[24] Straszewski T, Siegel JT. Positive Emotion Infusions: Can Savoring Increase Help-Seeking Intentions among People with Depression? Applied Psychology: Health and Well-Being. 2018;10(1):171-90. _eprint: https://onlinelibrary.wiley.com/doi/pdf/10.1111/aphw.12122. Available from: https://onlinelibrary.wiley.com/doi/abs/10.1111/aphw.12122.

[25] Ajzen I. The theory of planned behavior. Organizational Behavior and Human Decision Processes. 1991 Dec;50(2):179-211. Available from: https://www.sciencedirect.com/science/article/pii/074959789190020T.

[26] Breslin G, Shannon S, Prentice G, Rosato M, Leavey G. Adolescent Mental Health Help-Seeking from Family and Doctors: Applying the Theory of Planned Behaviour to the Northern Ireland Schools and Wellbeing Study. Child Care in Practice. 2022 Oct;28(4):522-35.

Publisher: Routledge _eprint: https://doi.org/10.1080/13575279.2021.1918639. Available from: https://doi.org/10.1080/13575279.2021.1918639.

[27] Adam A, Jain A, Pletnikova A, Bagga R, Vita A, N Richey L, et al. Use of a Mobile App to Augment Psychotherapy in a Community Psychiatric Clinic: Feasibility and Fidelity Trial. JMIR formative research. 2020 Jul;4(7):e17722.

[28] Liddle SK, Vella SA, Deane FP. Attitudes about mental illness and help seeking among adolescent males. Psychiatry Research. 2021 Jul;301:113965. Available from: https://www.sciencedirect.com/science/article/pii/S0165178121002626.

[29] Tomczyk S, Schomerus G, Stolzenburg S, Muehlan H, Schmidt S. Ready, Willing and Able? An Investigation of the Theory of Planned Behaviour in Help-Seeking for a Community Sample with Current Untreated Depressive Symptoms. Prevention Science. 2020 Aug;21(6):749-60. Available from: https://doi.org/10.1007/s11121-020-01099-2.

[30] Hussain SA, Alhabash S. Nostalgic Emotional Valence and Its Effects on Help-Seeking in Depression. An Application of the Theory of Planned Behavior. Health Communication. 2021 Nov;36(13):1731-42. Publisher: Routledge _eprint: https://doi.org/10.1080/10410236.2020.1794549. Available from: https://doi.org/10.1080/10410236.2020.1794549.

[31] Sanci L, Kauer S, Thuraisingam S, Davidson S, Duncan AM, Chondros P, et al. Effectiveness of a Mental Health Service Navigation Website (Link) for Young Adults: Randomized Controlled Trial. JMIR mental health. 2019 Oct;6(10):e13189.

[32] Hui A, Wong PWC, Fu KW. Evaluation of an Online Campaign for Promoting Help-Seeking Attitudes for Depression Using a Facebook Advertisement: An Online Randomized Controlled Experiment. JMIR Mental Health. 2015 Mar;2(1):e3649. Company: JMIR Mental Health Distributor: JMIR Mental Health Institution: JMIR Mental Health Label: JMIR Mental Health Publisher: JMIR Publications Inc., Toronto, Canada. Available from: https://mental.jmir.org/2015/1/e5.

[33] Deci EL, Eghrari H, Patrick BC, Leone DR. Facilitating Internalization: The Self-Determination Theory Perspective. Journal of Personality. 1994;62(1):119-42. _eprint: https://onlinelibrary.wiley.com/doi/pdf/10.1111/j.1467-6494.1994.tb00797.x. Available from: https://onlinelibrary.wiley.com/doi/abs/10.1111/j.1467-6494.1994.tb00797.x.

[34] Van Tiem J, Moeckli J, Suiter N, Fuhrmeister L, Pham K, Dindo L, et al. "A link to the outside:" Patient perspectives on a mobile texting program to improve depression self-management. Patient Education and Counseling. 2021 Sep;104(9):2154-8.

[35] Välimäki M, Kurki M, Hätönen H, Koivunen M, Selander M, Saarijärvi S, et al. Developing an internet-based support system for adolescents with depression. JMIR research protocols. 2012 Dec;1(2):e22.

[36] Alvarez-Jimenez M, Rice S, D'Alfonso S, Leicester S, Bendall S, Pryor I, et al. A Novel Multimodal Digital Service (Moderated Online Social Therapy+) for Help-Seeking Young People Experiencing Mental Ill-Health: Pilot Evaluation Within a National Youth E-Mental Health Service. Journal of Medical Internet Research. 2020 Aug;22(8):e17155.

[37] Suka M, Yamauchi T, Sugimori H. Help-seeking intentions for early signs of mental illness and their associated factors: comparison across four kinds of health problems. BMC Public Health. 2016 Apr;16(1):301. Available from: https://doi.org/10.1186/s12889-016-2998-9.

[38] Isaak CA, Mota N, Medved M, Katz LY, Elias B, Mignone J, et al. Conceptualizations of help-seeking for mental health concerns in First Nations communities in Canada: A comparison of fit with the Andersen Behavioral Model. Transcultural Psychiatry. 2020 Apr;57(2):346-62. Publisher: SAGE Publications Ltd. Available from: https://doi.org/10.1177/1363461520906978.

[39] Chong SA, Abdin E, Vaingankar JA, Kwok KW, Subramaniam M. Where do People with Mental Disorders in Singapore go to for Help? Annals of the Academy of Medicine, Singapore. 2012 Apr;41(4):154-60. Available from: https://annals.edu.sg/pdf/41VolNo4Apr2012/V41N4p154.pdf.

[40] Goldberg DP, Huxley P. Mental Illness in the Community: The Pathway to Psychiatric Care. Psychology Press; 2001.

[41] Giallo R, Dunning M, Gent A. Attitudinal barriers to help-seeking and preferences for mental health support among Australian fathers. Journal of Reproductive and Infant Psychology. 2017 May;35(3):236-47. Publisher: Routledge _eprint: https://doi.org/10.1080/02646838.2017.1298084. Available from: https://doi.org/10.1080/02646838.2017.1298084.

[42] Wiljer D, Shi J, Lo B, Sanches M, Hollenberg E, Johnson A, et al. Effects of a Mobile and Web App (Thought Spot) on Mental Health Help-Seeking Among College and University Students: Randomized Controlled Trial. Journal of Medical Internet Research. 2020 Oct;22(10):e20790.

[43] Shin YM, Jung HY, Kim SW, Lee SH, Shin SE, Park JI, et al. A descriptive study of pathways to care of high risk for psychosis adolescents in Korea. Early Intervention in Psychiatry. 2010;4(2):119-23. _eprint: https://onlinelibrary.wiley.com/doi/pdf/10.1111/j.1751-7893.2010.00180.x. Available from: https://onlinelibrary.wiley.com/doi/abs/10.1111/j.1751-7893.2010.00180.x.

[44] Davies B, Harré R. Positioning: The discursive production of selves. Journal for the theory of social behaviour. 1990;20(1):43-63. Publisher: Oxford. Available from: https://www.academia.edu/download/8651928/davids-harrc3a9-positioning-theory1.pdf.

[45] Prior S. Overcoming stigma: how young people position themselves as counselling service users. Sociology of Health & Illness. 2012;34(5):697-713. _eprint: https://onlinelibrary.wiley.com/doi/pdf/10.1111/j.1467-9566.2011.01430.x. Available from: https://onlinelibrary.wiley.com/doi/abs/10.1111/j.1467-9566.2011.01430.x.

[46] Tajfel H, Turner JC, Austin WG, Worchel S. An integrative theory of intergroup conflict. Organizational identity: A reader. 1979;56(65):9780203505984-16. Publisher: Oxford University Press. Available from: https://books.google.com/books?hl=en&lr=&id=BgBREAAAQBAJ&oi=fnd&pg=PA56&dq=Tajfel,+H.,+Turner,+J.+C.,+Austin,+W.+G.,+%26+Worchel,+S.+(1979).+An+integrative+theory+of+intergroup+conflict.+Organizational+identity:+A+reader,+56-65.&ots=5sRfLhkr6s&sig=MPUSCkVvgCrTtQllsMKHuv3t_u8.

[47] Kearns M, Muldoon OT, Msetfi RM, Surgenor PWG. Understanding help-seeking amongst university students: the role of group identity, stigma, and exposure to suicide and help-seeking. Frontiers in Psychology. 2015;6. Available from: https://www.frontiersin.org/articles/10.3389/fpsyg.2015.01462.

[48] Klik KA, Williams SL, Reynolds KJ. Toward understanding mental illness stigma and help-seeking: A social identity perspective. Social Science & Medicine. 2019 Feb;222:35-43. Available from: https://www.sciencedirect.com/science/article/pii/S0277953618306749.

[49] Saavedra J, Arias-Sánchez S, de la Mata PML, Matías-García JA. Social Positioning Analysis as a Qualitative Methodology to Study Identity Construction in People Diagnosed With Severe Mental Illnesses. Qualitative Health Research. 2022 Jan;32(2):360-70. Publisher: SAGE Publications Inc. Available from: https://doi.org/10.1177/10497323211050377.

[50] Major B, O'Brien LT. The Social Psychology of Stigma. Annual Review of Psychology. 2005;56(1):393-421. _eprint: https://doi.org/10.1146/annurev.psych.56.091103.070137. Available from: https://doi.org/10.1146/annurev.psych.56.091103.070137.

[51] Eisenberg D, Downs MF, Golberstein E, Zivin K. Stigma and Help Seeking for Mental Health Among College Students. Medical Care Research and Review. 2009 Oct;66(5):522-41. Publisher: SAGE Publications Inc. Available from: https://doi.org/10.1177/1077558709335173.

[52] Hall S, Cheston R. Mental health and identity: the evaluation of a drop-in centre. Journal of Community & Applied Social Psychology. 2002;12(1):30-43. _eprint: https://onlinelibrary.wiley.com/doi/pdf/10.1002/casp.639. Available from: https://onlinelibrary.wiley.com/doi/abs/10.1002/casp.639.

[53] Biddle L, Donovan J, Sharp D, Gunnell D. Explaining non-help-seeking amongst young adults with mental distress: a dynamic interpretive model of illness behaviour: Illness behaviour among young adults with mental distress. Sociology of Health & Illness. 2007 Aug;29(7):983-1002. Available from: https://onlinelibrary.wiley.com/doi/10.1111/j.1467-9566.2007.01030.x.

[54] Lynch L, Long M, Moorhead A. Young Men, Help-Seeking, and Mental Health Services: Exploring Barriers and Solutions. American Journal of Men's Health. 2018 Jan;12(1):138-49. Available from: https://doi.org/10.1177/1557988315619469.

[55] Gulliver A, Griffiths KM, Christensen H. Perceived barriers and facilitators to mental health help-seeking in young people: a systematic review. BMC Psychiatry. 2010 Dec;10(1):113. Available from: https://doi.org/10.1186/1471-244X-10-113.

[56] Rickwood DJ, Deane FP, Wilson CJ. When and how do young people seek professional help for mental health problems? Medical Journal of Australia. 2007;187(S7):S35-9. Available from: https://onlinelibrary.wiley.com/doi/abs/10.5694/j.1326-5377.2007.tb01334.x.

[57] Yamasaki S, Ando S, Shimodera S, Endo K, Okazaki Y, Asukai N, et al. The Recognition of Mental Illness, Schizophrenia Identification, and Help-Seeking from Friends in Late Adolescence. PLOS ONE. 2016 Mar;11(3):e0151298. Publisher: Public Library of Science. Available from: https://journals.plos.org/plosone/article?id=10.1371/journal.pone.0151298.

[58] Horsfield P, Stolzenburg S, Hahm S, Tomczyk S, Muehlan H, Schmidt S, et al. Self-labeling as having a mental or physical illness: the effects of stigma and implications for help-seeking. Social Psychiatry and Psychiatric Epidemiology. 2020 Jul;55(7):907-16. Available from: https://doi.org/10.1007/s00127-019-01787-7.

[59] O'Dea B, King C, Subotic-Kerry M, Anderson M, Achilles MR, Parker B, et al. Evaluating a Web-Based Mental Health Service for Secondary School Students in Australia: Protocol for a Cluster Randomized Controlled Trial. JMIR Research Protocols. 2019 May;8(5):e12892. Available from: https://www.researchprotocols.org/2019/5/e12892/.

[60] Pescosolido BA. Beyond Rational Choice: The Social Dynamics of How People Seek Help. American Journal of Sociology. 1992 Jan;97(4):1096-138. Available from: https://www.journals.uchicago.edu/doi/10.1086/229863.

[61] Beatie BE, Mackenzie CS, Thompson G, Koven L, Eschenwecker T, Walker JR. Exploring older adults' experiences seeking psychological services using the network episode model. Ageing & Society. 2022 Jan;42(1):48-71. Available from: https://www.cambridge.org/core/journals/ageing-and-society/article/exploring-older-adults-experiences-seeking-psychological-services-using-the-network-episode-model/6EF81BD26583F680499DD66BEC33A591.

[62] Boydell KM, Volpe T, Gladstone BM, Stasiulis E, Addington J. Youth at ultra high risk for psychosis: using the Revised Network Episode Model to examine pathways to mental health care. Early Intervention in Psychiatry. 2013;7(2):170-86. _eprint: https://onlinelibrary.wiley.com/doi/pdf/10.1111/j.1751-7893.2012.00350.x. Available from: https://onlinelibrary.wiley.com/doi/abs/10.1111/j.1751-7893.2012.00350.x.

[63] Arnault DS. Cultural Determinants of Help Seeking: A model for research and practice. Research and theory for nursing practice. 2009;23(4):259-78. Available from: https://www.ncbi.nlm.nih.gov/pmc/articles/PMC2796597/.

[64] Nohr L, Ruiz AL, Ferrer JES, Buhlmann U. Mental health stigma and professional help-seeking attitudes a comparison between Cuba and Germany. PLOS ONE. 2021 Feb;16(2):e0246501. Publisher: Public Library of Science. Available from: https://journals.plos.org/plosone/article?id=10.1371/journal.pone.0246501.

[65] Ohta R, Sato M, Kitayuguchi J, Maeno T, Sano C. Potential Help-Seeking Behaviors Associated with Better Self-Rated Health among Rural Older Patients: A Cross-Sectional Study. International Journal of Environmental Research and Public Health. 2021 Jan;18(17):9116. Number: 17 Publisher: Multidisciplinary Digital Publishing Institute. Available from: https://www.mdpi.com/1660-4601/18/17/9116.

[66] Campbell C, Burgess R. The role of communities in advancing the goals of the Movement for Global Mental Health. Transcultural Psychiatry. 2012 Jul;49(3-4):379-95. Publisher: SAGE Publications Ltd. Available from: https://doi.org/10.1177/1363461512454643.

[67] Mathias K, Mathias J, Goicolea I, Kermode M. Strengthening community mental health competence—A realist informed case study from Dehradun, North India. Health & Social Care in the Community. 2018;26(1):e179-90. _eprint: https://onlinelibrary.wiley.com/doi/pdf/10.1111/hsc.12498. Available from: https://onlinelibrary.wiley.com/doi/abs/10.1111/hsc.12498.

[68] Mannell J, Seyed-Raeisy I, Burgess R, Campbell C. The implications of community responses to intimate partner violence in Rwanda. PLOS ONE. 2018 May;13(5):e0196584. Publisher: Public Library of Science. Available from: https://journals.plos.org/plosone/article?id=10.1371/journal.pone.0196584.

[69] Pillai P, Rawat M, Jain S, Martin RA, Shelly K, Mathias K. Developing relevant community mental health programmes in North India: five questions we ask when co-producing knowledge with experts by experience. BMJ Global Health. 2023 Aug;8(8):e011671. Publisher: BMJ Specialist Journals Section: Practice. Available from: https://gh.bmj.com/content/8/8/e011671.

[70] Murnane EL, Walker TG, Tench B, Voida S, Snyder J. Personal Informatics in Interpersonal Contexts: Towards the Design of Technology that Supports the Social Ecologies of Long-Term Mental Health Management. Proceedings of the ACM on Human-Computer Interaction. 2018 Nov;2(CSCW):127:1-127:27. Available from: https://dl.acm.org/doi/10.1145/3274396.

[71] Marcu G, Huh-Yoo J. Attachment-Informed Design: Digital Interventions That Build Self-Worth, Relationships, and Community in Support of Mental Health. In: Proceedings of the 2023 ACM Designing Interactive Systems Conference. DIS '23. New York, NY, USA: Association for Computing Machinery; 2023. p. 2453-68. Available from: https://doi.org/10.1145/3563657.3596009.

[72] Adams C, Gringart E, Strobel N. Explaining adults' mental health help-seeking through the lens of the theory of planned behavior: a scoping review. Systematic Reviews. 2022 Aug;11(1):160. Available from: https://doi.org/10.1186/s13643-022-02034-y.

[73] Bhui K, Bhugra D. Mental illness in Black and Asian ethnic minorities: pathways to care and outcomes. Advances in Psychiatric Treatment. 2002 Jan;8(1):26-33. Available from: https://www.cambridge.org/core/journals/advances-in-psychiatric-treatment/article/mental-illness-in-black-and-asian-ethnic-minorities-pathways-to-care-and-outcomes/B44BECFA32EBBABCA9105E2792F92BF5.

[74] Gater R, Almeida E Sousa DB, Barrientos G, Caraveo J, Chandrashekar CR, Dhadphale M, et al. The pathways to psychiatric care: a cross-cultural study. Psychological Medicine. 1991 Aug;21(3):761-74. Available from: https://www.cambridge.org/core/product/identifier/S003329170002239X/type/journal_article.

[75] Pendse SR, Sharma A, Vashistha A, De Choudhury M, Kumar N. "Can I Not Be Suicidal on a Sunday?": Understanding Technology-Mediated Pathways to Mental Health Support. In: Proceedings of the 2021 CHI Conference on Human Factors in Computing Systems. CHI '21. New York, NY, USA: Association for Computing Machinery; 2021. p. 1-16. Available from: https://doi.org/10.1145/3411764.3445410.

[76] Andersen R. A behavioral model of families' use of health services. A behavioral model of families' use of health services. 1968;(25). Available from: https://www.cabdirect.org/cabdirect/abstract/19702701913.

[77] Andersen RM. Revisiting the Behavioral Model and Access to Medical Care: Does it Matter? Journal of Health and Social Behavior. 1995 Mar;36(1):1. Available from: http://www.jstor.org/stable/2137284?origin=crossref.

[78] Babitsch B, Gohl D, von Lengerke T. Re-revisiting Andersen's Behavioral Model of Health Services Use: a systematic review of studies from 1998–2011. GMS Psycho-Social-Medicine. 2012 Oct;9:Doc11. Available from: https://www.ncbi.nlm.nih.gov/pmc/articles/PMC3488807/.

[79] Dong X, Simon M, Beck T, Evans D. A cross-sectional population-based study of elder self-neglect and psychological, health, and social factors in a biracial community. Aging & Mental Health. 2010 Jan;14(1):74-84. Available from: https://doi.org/10.1080/13607860903421037.

[80] Lu W, Todhunter-Reid A, Mitsdarffer ML, Muñoz-Laboy M, Yoon AS, Xu L. Barriers and Facilitators for Mental Health Service Use Among Racial/Ethnic Minority Adolescents: A Systematic Review of Literature. Frontiers in Public Health. 2021 Mar;9:641605. Available from: https://www.frontiersin.org/articles/10.3389/fpubh.2021.641605/full.

[81] Stets JE, Burke PJ. Identity Theory and Social Identity Theory. Social Psychology Quarterly. 2000;63(3):224-37. Available from: https://www.jstor.org/stable/2695870.

[82] Martínez-Hernáez A, DiGiacomo SM, Carceller-Maicas N, Correa-Urquiza M, Martorell-Poveda MA. Non-professional-help-seeking among young people with depression: a qualitative study. BMC Psychiatry. 2014 Apr;14(1):124. Available from: https://doi.org/10.1186/1471-244X-14-124.

[83] Abavi R, Branston A, Mason R, Du Mont J. An Exploration of Sexual Assault Survivors' Discourse Online on Help-Seeking. Violence and Victims. 2020 Feb;35(1):126-40.

[84] Lannin DG, Vogel DL, Brenner RE, Abraham WT, Heath PJ. Does self-stigma reduce the probability of seeking mental health information? Journal of Counseling Psychology. 2016 Apr;63(3):351-8. Available from: http://doi.apa.org/getdoi.cfm?doi=10.1037/cou0000108.

[85] Pescosolido BA. Organizing the Sociological Landscape for the Next Decades of Health and Health Care Research: The Network Episode Model III-R as Cartographic Subfield Guide. In: Pescosolido BA, Martin JK, McLeod JD, Rogers A, editors. Handbook of the Sociology of Health, Illness, and Healing: A Blueprint for the 21st Century. Handbooks of Sociology and Social Research. New York, NY: Springer; 2011. p. 39-66. Available from: https://doi.org/10.1007/978-1-4419-7261-3_3.

[86] Campbell R, Dworkin E, Cabral G. An Ecological Model of the Impact of Sexual Assault On Women's Mental Health. Trauma, Violence, & Abuse. 2009 Jul;10(3):225-46. Available from: https://doi.org/10.1177/1524838009334456.

[87] Brooks HL, Bee P, Lovell K, Rogers A. Negotiating support from relationships and resources: a longitudinal study examining the role of personal support networks in the management of severe and enduring mental health problems. BMC Psychiatry. 2020 Dec;20(1):50. Available from: https://bmcpsychiatry.biomedcentral.com/articles/10.1186/s12888-020-2458-z.

[88] Shahid M, Weiss NH, Stoner G, Dewsbury B. Asian Americans' mental health help-seeking attitudes: The relative and unique roles of cultural values and ethnic identity. Asian American Journal of Psychology. 2021;12:138-46.

[89] Kruzan KP, Fitzsimmons-Craft EE, Dobias M, Schleider JL, Pratap A. Developing, Deploying, and Evaluating Digital Mental Health Interventions in Spaces of Online Help- and Information-Seeking. Procedia Computer Science. 2022 Jan;206:6-22. Available from: https://www.sciencedirect.com/science/article/pii/S1877050922009541.

[90] Kaziunas E, Klinkman MS, Ackerman MS. Precarious Interventions: Designing for Ecologies of Care. Proceedings of the ACM on Human-Computer Interaction. 2019 Nov;3(CSCW):113:1-113:27. Available from: https://dl.acm.org/doi/10.1145/3359215.

[91] Ojio Y, Mori R, Matsumoto K, Nemoto T, Sumiyoshi T, Fujita H, et al. Innovative approach to adolescent mental health in Japan: School-based education about mental health literacy. Early Intervention in Psychiatry. 2021 Feb;15(1):174-82.

[92] Migliorini C, Barrington N, O'Hanlon B, O'Loughlin G, Harvey C. The Help-Seeking Experiences of Family and Friends Who Support Young People With Mental Health Issues: A Qualitative Study. Qualitative Health Research. 2023 Feb;33(3):191-203.

[93] Mueller NE, Panch T, Macias C, Cohen BM, Ongur D, Baker JT. Using Smartphone Apps to Promote Psychiatric Rehabilitation in a Peer-Led Community Support Program: Pilot Study. JMIR mental health. 2018 Aug;5(3):e10092.

[94] Pelttari S. Emotional self-disclosure and stance-taking within affective narratives on YouTube: A qualitative case study of four Spanish YouTubers. International Journal of Language and Culture. 2022 Dec;9(2):292-321. Available from: https://www.jbe-platform.com/content/journals/10.1075/ijolc.22050.pel.

[95] Tan Y, Lattie EG, Qiu Y, Teng Z, Wu C, Tang H, et al. Accessibility of mental health support in China and preferences on web-based services for mood disorders: A qualitative study. Internet Interventions. 2021 Dec;26:100475. Available from: https://www.sciencedirect.com/science/article/pii/S2214782921001159.

[96] Pendse SR, Karusala N, Siddarth D, Gonsalves P, Mehrotra S, Naslund JA, et al. Mental health in the global south: challenges and opportunities in HCI for development. In: Proceedings of the 2nd ACM SIGCAS Conference on Computing and Sustainable Societies. COMPASS '19. New York, NY, USA: Association for Computing Machinery; 2019. p. 22-36. Available from: https://dl.acm.org/doi/10.1145/3314344.3332483.

[97] Li G, Zhou X, Lu T, Yang J, Gu N. SunForum: Understanding Depression in a Chinese Online Community. In: Proceedings of the 19th ACM Conference on Computer-Supported Cooperative Work & Social Computing. CSCW '16. New York, NY, USA: Association for Computing Machinery; 2016. p. 515-26. Available from: https://doi.org/10.1145/2818048.2819994.

[98] Jones HJ, Norwood CR, Bankston K. Leveraging Community Engagement to Develop Culturally Tailored Stress Management Interventions in Midlife Black Women. Journal of Psychosocial Nursing and Mental Health Services. 2019 Mar;57(3):32-8. Publisher: SLACK Incorporated. Available from: https://journals.healio.com/doi/10.3928/02793695-20180925-01.

[99] Boyd JE, Adler EP, Otilingam PG, Peters T. Internalized Stigma of Mental Illness (ISMI) scale: a multinational review. Comprehensive psychiatry. 2014;55(1):221-31.

[100] Williams KDA, Wijaya C, Stamatis CA, Abbott G, Lattie EG. Insights Into Needs and Preferences for Mental Health Support on Social Media and Through Mobile Apps Among Black Male University Students: Exploratory Qualitative Study. JMIR formative research. 2022 Aug;6(8):e38716.

[101] Chen Y. Exploring the Effect of Social Support and Empathy on User Engagement in Online Mental Health Communities. International Journal of Environmental Research and Public Health. 2021;18(13):6855. Available from: https://www.proquest.com/docview/2549337448/abstract/185F145ABDEB4BABPQ/1?parentSessionId=SVZ7a7isTJrfPhEMByzhQz2TzKi73h8I%2FIvuuhvdvWk%3D&sourcetype=Scholarly%20Journals.

[102] Daniel M, Maulik PK, Kallakuri S, Kaur A, Devarapalli S, Mukherjee A, et al. An integrated community and primary healthcare worker intervention to reduce stigma and improve management of common mental disorders in rural India: protocol for the SMART Mental Health programme. Trials. 2021 Mar;22(1):179.

[103] Alqahtani F, Winn A, Orji R. Co-Designing a Mobile App to Improve Mental Health and Well-Being: Focus Group Study. JMIR Formative Research. 2021 Feb;5(2):e18172. Available from: https://formative.jmir.org/2021/2/e18172.

[104] Davis FD, Bagozzi R, Warshaw P. Technology acceptance model. J Manag Sci. 1989;35(8):982-1003.

[105] Mayer G, Hummel S, Oetjen N, Gronewold N, Bubolz S, Blankenhagel K, et al. User experience and acceptance of patients and healthy adults testing a personalized self-management app for depression: A non-randomized mixed-methods feasibility study. Digital Health. 2022;8:20552076221091353.

[106] Burgess ER, Zhang R, Ernala SK, Feuston JL, De Choudhury M, Czerwinski M, et al. Technology ecosystems: rethinking resources for mental health. Interactions. 2021 Jan;28(1):66-71. Available from: https://dl.acm.org/doi/10.1145/3434564.

[107] Liu J, Zhang Y. Understanding and Facilitating Mental Health Help-Seeking of Young Adults: A Socio-technical Ecosystem Framework. arXiv; 2024. Available from: http://arxiv.org/abs/2401.08994.

[108] Zhang R, E Ringland K, Paan M, C Mohr D, Reddy M. Designing for Emotional Well-being: Integrating Persuasion and Customization into Mental Health Technologies. In: Proceedings of the 2021 CHI Conference on Human Factors in Computing Systems. Yokohama Japan: ACM; 2021. p. 1-13. Available from: https://dl.acm.org/doi/10.1145/3411764.3445771.

[109] Trainum K, Liu J, Hauser E, Xie B. Nursing Staff's Attitudes, Needs, and Preferences for Care Robots in Assisted Living Facilities: A Systematic Literature Review. In: Companion of the 2024 ACM/IEEE International Conference on Human-Robot Interaction; 2024. p. 1058-62.

[110] Pinto MD, Hickman RL, Clochesy J, Buchner M. Avatar-based depression self-management technology: promising approach to improve depressive symptoms among young adults...Self-Management Resource Training for Mental Health (eSMART-MH). Applied Nursing Research. 2013 Feb;26(1):45-8.